\documentclass[11pt,a4paper]{article}
\usepackage{amstext,graphicx}

\newcommand{\Tr}{\mathop{\mathrm{Tr}}\nolimits}
\newcommand{\E}{\mathop{\mathrm{Ein}}\nolimits}
\newcommand{\tfrac}[2]{{\textstyle\frac{#1}{#2}}}

\begin{document}

\begin{titlepage}
\begin{flushright}
WUE/ITP--97--016\\
hep-ph/9706421
\end{flushright}
\vspace{1cm}
\begin{center}
\huge Couplings of heavy hadrons with soft pions\\
from QCD sum rules\\[1cm]
\normalsize A.~G.~Grozin$^*$\\
\textsl{Institut f\"ur Physik, Johannes--Gutenberg--Universit\"at,}\\
\textsl{Staudinger Weg 7, D--55099 Mainz, Germany}\\
O.~I.~Yakovlev$^*$\\
\textsl{Institut f\"ur Theoretische Physik II,
Bayrische--Maximilians--Universit\"at},\\
\textsl{Am Hubland, D--97074 W\"urzburg, Germany}
\end{center}

\begin{abstract}
We estimate the couplings in the Heavy Hadron Chiral Theory (HHCT) lagrangian
from the QCD sum rules in an external axial field.
We take into account the perturbative correction to the meson correlator
in the infinite mass limit.
With the perturbative correction and three successive power corrections,
the meson correlator in an axial field
becomes one of the best known correlators.
In spite of this, the corresponding sum rule is not very stable.
It yields the result $g_1 F^2/(380\text{MeV})^3=0.1\div 0.2$,
where $F^2=f^2_M m/4=(380\text{MeV})^3$ is the central value
of the heavy meson decay constant with the perturbative correction~\cite{BBBD}.
This result is surprisingly low as compared with the constituent quark
model estimate $g_1=0.75$.
The sum rules for $g_{2,3}$ following from nondiagonal 
$\Sigma-\Sigma$ and diagonal $\Lambda-\Sigma$ baryon correlators in
an external axial field suggest 
$g_{2,3}= 0.4\div 0.7$,
while diagonal
$\Sigma-\Sigma$ and nondiagonal $\Lambda-\Sigma$ baryon sum rules
have too large uncertainties.
\end{abstract}

\vfill

\noindent
$^*$ On leave from Budker Institute of Nuclear Physics,
Novosibirsk 630090, Russia.
\end{titlepage}

\section{Introduction}
\label{Intro}

It is well known that the QCD lagrangian with $n_l$ massless flavours
has the $SU(n_l)_L\times SU(n_l)_R$ symmetry spontaneously broken to $SU(n_l)_V$
giving the $(n_l^2-1)$--plet of pseudoscalar massless Goldstone mesons (pions)
$\pi^i_j$ ($\pi^i_i=0$).
Their interactions at low momenta are described by the chiral lagrangian
(see e.~g.~\cite{Georgi})
\begin{equation}
L_\pi = \frac{f_\pi^2}{8}\Tr \partial_\mu \Sigma^+ \partial^\mu \Sigma
+ \cdots,
\quad
\Sigma = \exp \frac{2i\pi}{f_\pi},
\label{Lpi}
\end{equation}
where the pion constant $f_\pi\approx132$MeV is defined by
\[
{<}0|j^i_{j\mu}|\pi{>} = if_\pi e^i_j p_\mu, \quad
j^i_{j\mu} = \overline{q}_j \gamma_\mu\gamma_5 q^i \to
i\frac{f_\pi^2}{2} \left(\Sigma^+\partial_\mu \Sigma
- \Sigma\partial_\mu \Sigma^+\right)^i_j
\]
($e^i_j$ is the pion flavour wave function),
and dots mean terms with more derivatives.
Light quark masses can be included perturbatively,
and lead to extra terms in~(\ref{Lpi}).
$SU(n_l)_L\times SU(n_l)_R$ transformations act as $\Sigma \to L\Sigma R^+$.
Let's define $\xi=\exp i\pi/f_\pi$, $\Sigma=\xi^2$;
it transforms as $\xi \to L\xi U^+ = U\xi R^+$
where $U$ is a $SU(n_l)$ matrix depending on $\pi(x)$.
The vector $v_\mu=\frac{1}{2}(\xi^+\partial_\mu \xi+\xi\partial_\mu \xi^+)$
and the axial vector
$a_\mu=\frac{i}{2}(\xi^+\partial_\mu \xi-\xi\partial_\mu \xi^+)$
transform as $v_\mu \to U(v_\mu+\partial_\mu)U^+$, $a_\mu \to U a_\mu U^+$.
There is a freedom in transformation laws of matter fields such as $\psi^i$
because it is always possible to multiply them by a matrix depending on $\pi$.
The only requirement is the correct transformation with respect to $SU(n_l)_V$
($L=R$).
It is convenient to choose $\psi\to U\psi$.
Then the covariant derivative $D_\mu=\partial_\mu+v_\mu$
transforms as $D_\mu\psi\to UD_\mu\psi$.
Covariant derivatives of tensors with more flavour indices
are defined similarly.

Hadrons with a heavy quark are now successively investigated
in the framework of the Heavy Quark Effective Theory (HQET)
(see~\cite{Neubert} for review and references).
To the leading order in $1/m$, the heavy quark spin does not interact
and can be rotated or switched off at all
(spin--flavour and superflavour symmetry).
The $\overline{Q}q$ mesons with a spinless heavy quark
form the $j^P=\frac{1}{2}^+$ $n_l$--plet $\psi^i$.
The $Qqq$ baryons can have $j^P=0^+$ or $1^+$
giving the scalar flavour--antisymmetric $n_l(n_l-1)/2$--plet $\Lambda^{ik}$
and the vector flavour--symmetric $n_l(n_l+1)/2$--plet $\vec{\Sigma}\,^{ik}$.
Switching the heavy quark spin on
gives degenerate $0^-$ and $1^-$ $B$ and $B^*$ mesons,
$\frac{1}{2}^+$ $\Lambda$ baryons,
 and degenerate $\frac{1}{2}^+$ and $\frac{3}{2}^+$
$\Sigma$ and $\Sigma^*$ baryons.

Interaction of these ground--state heavy hadrons with soft pions
is described by the Heavy Hadron Chiral Theory~\cite{HHCT,Rad}
(see~\cite{Wise,Gatto} for review and references).
We start from the HHCT lagrangian with the heavy quark spin
switched off~\cite{We}:
\begin{eqnarray}
&&L = L_\pi + \overline{\psi}_i iD_0 \psi^i
  + \Lambda^*_{ij} iD_0 \Lambda^{ij}
  + \vec{\Sigma}\,^*_{ij} \cdot (iD_0-\Delta) \vec{\Sigma}\,^{ij}
\label{LHHCT}\\
&& + g_1 \overline{\psi}_i \rlap{/}{a}\,^i_j \gamma_5 \psi^j
  + 2ig_2 \vec{\Sigma}\,^*_{ik} \cdot \vec{a}\,^i_j \times \vec{\Sigma}\,^{jk}
  + 2g_3 \left( \Lambda^*_{ik} \vec{a}\,^i_j \cdot \vec{\Sigma}\,^{jk}
        + \vec{\Sigma}\,^*_{ik} \cdot \vec{a}\,^i_j \Lambda^{jk} \right),
\nonumber
\end{eqnarray}
where $\Delta$ is the $\Sigma$--$\Lambda$ mass difference.
The possibility of consideration of the $\Sigma\Lambda\pi$ interaction
in HHCT relies on the fact that this difference is small
compared to the chiral symmetry breaking scale
though formally both of them are of the order of the characteristic
hadron mass scale.
The matrix elements of the axial current
between heavy hadrons are easily obtained using PCAC:
\begin{eqnarray}
&&{<}M'|\vec{\jmath}\,^i_j|M{>} = g_1 \overline{u}'_j \vec{\gamma}\gamma_5 u^i,
\label{PCAC}\\
&&{<}\Sigma'|\vec{\jmath}\,^i_j|\Sigma{>}
 = 2ig_2 \vec{e}\,^{\prime*}_{jk} \times \vec{e}\,^{ik}, \quad
{<}\Lambda|\vec{\jmath}\,^i_j|\Sigma{>}
 = 2g_3 e^*_{jk} \vec{e}\,^{ik},
\nonumber
\end{eqnarray}
where $u^i$, $e^{ij}$, $\vec{e}\,^{ij}$ are the $M$, $\Lambda$, $\Sigma$
wave functions, and the nonrelativistic normalization of the states and
wave functions is assumed.
If we switch the heavy quark spin on,
we obtain the usual HHCT lagrangian~\cite{HHCT}.

The HHCT couplings $g_i$ should be in principle calculable in the underlying
theory---HQET, but this is a difficult nonperturbative problem.
Some experimental information is available on $g_1$ and $g_3$.
If we neglect $1/m_c$ corrections, then
$\Gamma(D^{*+}\to D^0\pi^+) = g_1^2 p_\pi^3/(6\pi f_\pi^2)$,
and similarly for $D^+\pi^0$ (with the extra $1/2$).
The experimental upper limit $\Gamma(D^{*+})<0.131$MeV
combined with the branching ratio $B(D^{*+}\to D^0\pi^+)=68\%$~\cite{PDG}
gives $g_1<0.68$.
A combined analysis of $D^*$ pionic and radiative decays~\cite{Rad}
gives $g_1\sim0.4$--$0.7$.
The recent CLEO measurement~\cite{CLEO} of the $\Sigma_c^*\to\Lambda_c\pi$
decays gives~\cite{gbar} $g_3=0.7\pm0.1$.
The decay $\Sigma_c^*\to\Sigma_c\pi$ is kinematically forbidden;
therefore, the direct measurement of $g_2$ is not possible.

In the constituent quark model, $g_1$ is the axial charge $g$
of the constituent light quark in the heavy meson.
Moreover, following the folklore definition
``constituent quark is $B$ meson minus $b$ quark'',
this is the most clear way to define $g$ of the constituent quark.
The baryonic couplings $g_{2,3}$ are also equal to $g$ in this model.
The most naive estimate is $g\approx1$;
the nucleon axial charge is $g_A=\frac{5}{3}g$ in the constituent model,
and in order to obtain $g_A=1.25$ we should assume $g=0.75$.

Sum rules~\cite{SVZ} were successfully used
to solve many nonperturbative problems in QCD and HQET.
The currents with the quantum numbers of the ground state mesons and baryons
with the heavy quark spin switched off are
\begin{equation}
j^i_M=Q^*\frac{1+\gamma_0}{2}q^i, \quad
j^{ij}_{\Lambda1}=(q^{T[i}C\gamma_5 q^{j]})Q, \quad
\vec{\jmath}\,^{ij}_{\Sigma1}=(q^{T(i}C\vec{\gamma}q^{j)})Q,
\label{Curr}
\end{equation}
where $Q$ is the spinless static quark field,
$C$ is the charge conjugation matrix, $q^T$ means $q$ transposed,
$(ij)$ and $[ij]$ mean symmetrization and antisymmetrization.
There are also currents $j^{ij}_{\Lambda2}$ and $\vec{\jmath}\,^{ij}_{\Sigma2}$
with the additional $\gamma_0$.
Correlator of the mesonic currents was investigated
in~\cite{Shuryak,BG,BBBD},
and of the baryonic ones---in~\cite{GY,GKY}.
The sum rules results are in a qualitative agreement
with the constituent quark model: the massless quark propagator
plus the quark condensate contribution
simulate the constituent quark propagation well enough.

The sum rules method was generalized to the case of a constant external field
for calculation of such static characteristics of hadrons
as the magnetic moments~\cite{IS}.
Sum rules in an external axial field were used~\cite{BK}
for calculation of $g_A$ of light baryons.
In the present work we use HQET sum rules in an axial field
to estimate $g_{1,2,3}$.

We introduce the external axial field $A^i_{j\mu}$ ($A^i_{i\mu}=0$)
by adding the term
\begin{equation}
\Delta L=j^{i\mu}_j A^j_{i\mu}
\label{Field}
\end{equation}
to the lagrangian.
We are going to calculate correlators of the currents~(\ref{Curr})
up to the terms linear in $A$ (these terms are denoted by the subscript $A$).
The light quark propagator in the gauge $x_\mu A_\mu(x)=0$
gets the contribution (whose first term was found in~\cite{BK})
\begin{equation}
S^i_{jA}(x,0)=-\frac{iA^i_j\cdot x}{2\pi^2}
\left(\frac{\rlap{/}{x}\gamma_5}{x^4}
-\frac{x^\mu\tilde{G}_{\mu\nu}\gamma^\nu}{4x^2}\right)+\cdots
\label{Prop}
\end{equation}
The $G^2$ term in $S_A$ vanishes after the vacuum averaging; we are not
going to calculate gluonic contributions beyond $G^2$ and hence may omit
this term.
The axial field induces the quark condensates
\begin{eqnarray}\label{inducedqq}
&&{<}q^{ia\alpha}(x)\overline{q}_{jb\beta}(0){>}_A = \frac{\delta^a_b}{4N}
\Bigl\{\Bigl[ f^2\rlap{/}{A}\,^i_j
+ \frac{i}{6}{<}\overline{q}q{>} [\rlap{/}{A}\,^i_j,\rlap{/}{x}]
\nonumber\\
&&\quad{} + \frac{m_1^2 f_\pi^2}{36} \left( 5 x^2 \rlap{/}{A}\,^i_j
- 2 A^i_j\cdot x\rlap{/}{x} \right) \Bigr] \gamma_5 \Bigr\}^\alpha_\beta,
\label{Qcond}\\
&&{<}q^{ia\alpha}g\tilde{G}\,^A_{\mu\nu}\overline{q}_{jb\beta}{>}_A =
- \frac{\left(t^A\right)^a_b}{12C_F N} m_1^2 f_\pi^2
\left(A^i_{j\mu}\gamma_\nu-A^i_{j\nu}\gamma_\mu \right)^\alpha_\beta,
\nonumber
\end{eqnarray}
where $m_1^2\approx0.2\text{GeV}^2$~\cite{Use,CZZ} is defined by
${<}0|\overline{d}g\tilde{G}_{\mu\nu}\gamma^\nu u|\pi^+{>}
=im_1^2f_\pi p_\mu$, $N=3$ is the number of colours, $C_F=(N^2-1)/(2N)$
(the sum rule considered in~\cite{CZZ} yields the relation
$m_1^2\approx m_0^2/4$).
This formula has been first obtained in~\cite{BK}, with the wrong sign
of $m_1^2$ terms (this sign was changed in~\cite{BIK},
where a flavour--singlet external field was considered).
We assume $p\cdot A=0$ where $p\to0$ is the momentum of the field $A$;
in general these formulae should contain $A_\bot=A-(A\cdot p/p^2)p$.

The $D^*D\pi$ coupling was estimated~\cite{EK} from the sum rules
in an external axial field at a finite $m_c$ with the accuracy
up to $m_1^2$ term.
This last term is incorrect; however, it is rather small.
The large $m_c$ limit was analyzed in~\cite{Ovch,We}.
We think that the treatment of the phenomenological side
of the sum rule in~\cite{Ovch} is not quite convincing
(see Sec.~\ref{Mes}).
In our previous work~\cite{We}, we used an incorrect formula for the induced
condensates, which leads to the wrong sign of the $m_1^2$ term.
After correcting this, the result agrees with~\cite{Ovch,Ital,BBKR}.
In~\cite{Ital}, the two--current product sandwiched between the vacuum
and a soft pion was considered; this is equivalent to the two--point
sum rules in an external axial field.
Both finite and infinite mass case are discussed in this work,
with the extra power correction $m_0^2{<}\overline{q}q{>}$ included
(it is rather small).
Light--cone sum rules for both finite and infinite quark mass
were considered in~\cite{BBKR} (see also~\cite{CF}).
In this approach, the leading term of the correlator and power corrections
are determined by the leading-- and subleading--twist pion wave functions 
at $x=1/2$.
These values were determined from QCD sum rules~\cite{BF},
and are not so accuratly known as the parameters of the 
external--field sum rules, $f_\pi$ and ${<}\overline{q}q{>}$.
In the soft pion limit, the formulae of~\cite{Ital} are reproduced
(without the highest power correction).
The $D^*D\pi$ coupling (at both finite and infinite quark mass) was also
considered in the framework of 3--point QCD sum rules,
double Borel~\cite{Aliev} and double moments~\cite{DN}.
It is difficult to compare this approach with the one considered here,
because there correlators are determined by a completely different set
of contributions (with even dimensions).

Perturbative corrections to sum rules were not discussed in the above works.
Here we present the next--to--leading perturbative term in the infinite
mass limit.
With this term taken into account, the correlator used in the $g_1$ sum rule
is one of the best known correlators:
in addition to the leading term, one perturbative correction
and three successive power corrections are known.
In spite of this, the sum rule is not very stable (Sec.~\ref{Mes}).
We also discuss sum rules for the baryonic couplings $g_{2,3}$
(Sec.~\ref{Bar}), following~\cite{We} with some corrections.
The results for diagonal $\Lambda$--$\Sigma$ and
nondiagonal $\Sigma$--$\Sigma $ correlators suggest $g_{2,3}=0.4\div 0.7$.
The sum rules based on  nondiagonal $\Lambda$--$\Sigma$ and
diagonal $\Sigma$--$\Sigma$ correlators have strong Borel parameter
dependence and therefore yield no definite predictions.

\section{Mesons}
\label{Mes}

\begin{figure}[ht]
\includegraphics[width=\linewidth]{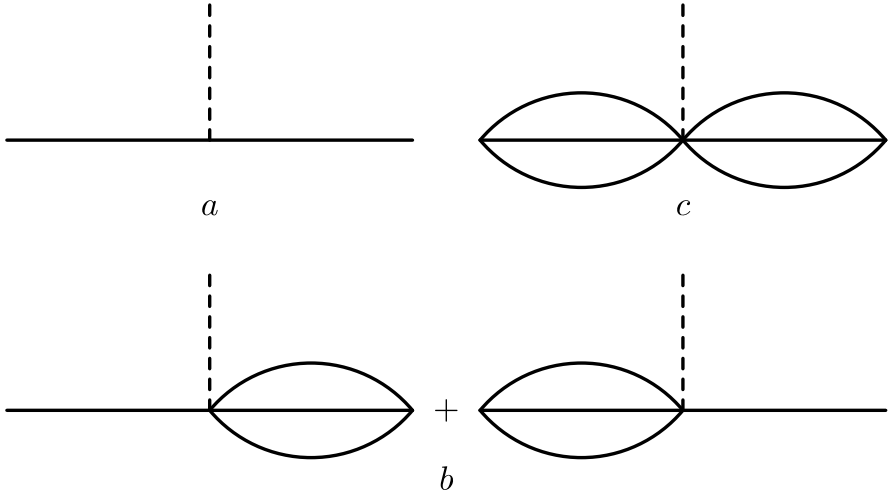}
\caption{Physical states' contributions to the correlator.}
\label{Fphys}
\end{figure}

The correlator of the meson currents~(\ref{Curr}) has the $A$--term
\begin{equation}
i{<}T j^i(x) \overline{\jmath}_j(0){>}_A
= \frac{1+\gamma_0}{2} \rlap{/}{A}\,^i_j
\gamma_5 \frac{1+\gamma_0}{2} \delta(\vec{x}) \Pi_A(x_0)
\label{Mcor}
\end{equation}
that depends only on $\vec{A}\,^i_j$ and not $A^i_{j0}$.
The correlator possesses the usual dispersion representation
at any $A^i_{j\mu}$.
The meson contribution at $A^i_{j\mu}=0$ is
$\rho(\omega)=F^2\delta(\omega-\varepsilon)$
where $\varepsilon$ is the meson energy and ${<}0|j^i|M{>}=iFu^i$
(the usual meson constant is $f_M=2F/\sqrt{m}$).
Switching on the external field produces the energy shift
$\varepsilon\to\varepsilon-g_1\gamma_5\vec\gamma\cdot\vec{A}$
($s_n=\frac{1}{2}\gamma_5\vec\gamma\cdot\vec{n}=\pm\frac{1}{2}$
is the meson spin projection).
Therefore the meson contribution is
$\rho_A(\omega)=g_1F^2\delta'(\omega-\varepsilon)+c\delta(\omega-\varepsilon)$
where the second term originates from the change of $F^2$.
Besides that there is a smooth continuum contribution
$\rho_A^{\text{cont}}(\omega)$.
Thus we obtain
\begin{equation}
\Pi_A(\omega)=\frac{g_1 F^2}{(\omega-\varepsilon)^2}
+\frac{c}{\omega-\varepsilon}+\Pi_A^{\text{cont}}(\omega).
\label{Phys}
\end{equation}
In other words, the lowest meson's contribution in both channels
(Fig.~\ref{Fphys}a) gives the double pole at $\omega=\varepsilon$;
mixed lowest--higher state contributions (Fig.~\ref{Fphys}b) give a single
pole at $\omega=\varepsilon$ plus a term with a spectral density in the
continuum region after the partial fraction decomposition;
higher states' contributions (Fig.~\ref{Fphys}c) have a spectral density
in the continuum region only.

\begin{figure}[p]
\includegraphics[width=\linewidth]{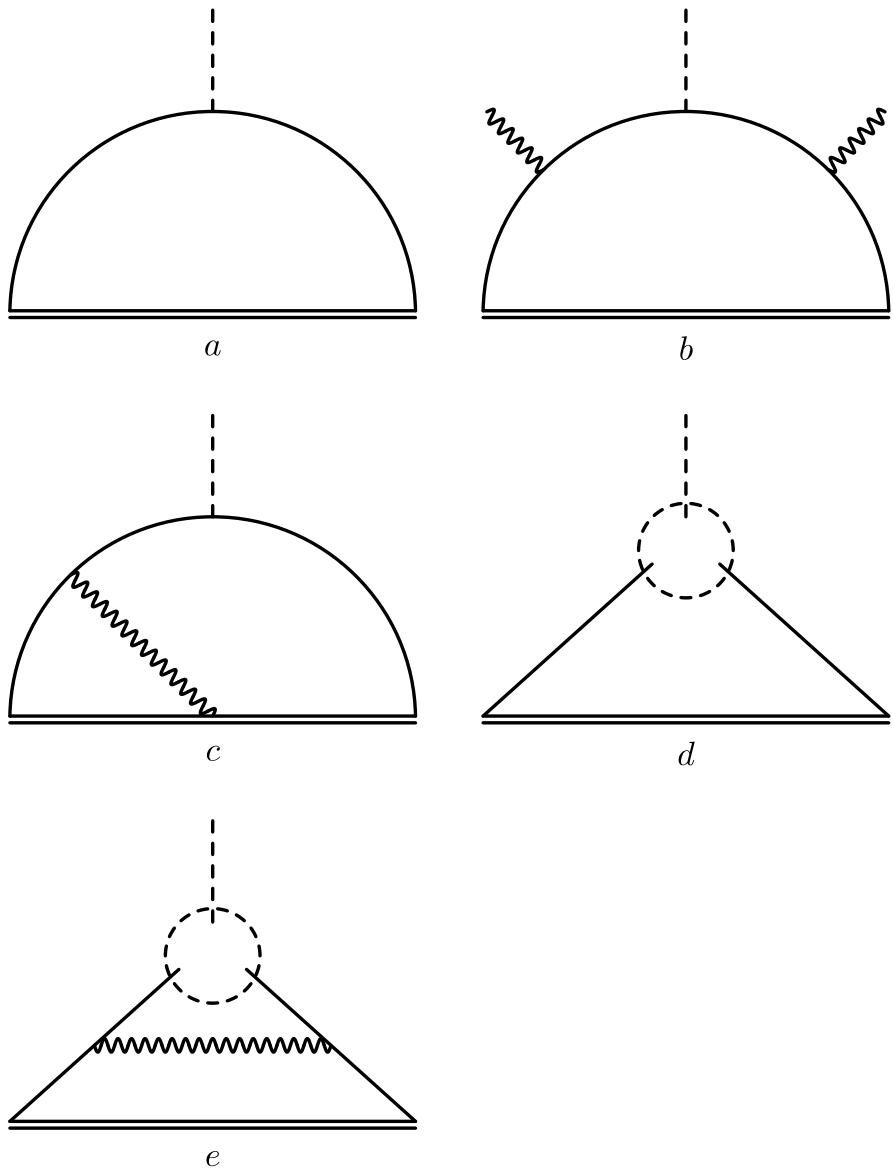}
\caption{Correlator of meson currents.}
\label{Fmes}
\end{figure}

We can also calculate the correlator using OPE.
Gluons don't interact with the heavy quark
in the fixed point gauge $x_\mu A_\mu(x)=0$.
The leading perturbative diagram (Fig.~\ref{Fmes}a) vanishes.
In fact, all diagrams with non--cut light quark line,
including gluon condensate contributions (Fig.~\ref{Fmes}b)
and higher loop corrections (Fig.~\ref{Fmes}c), vanish.
The reason is that we can anticommute $\gamma_5$ out,
and obtain a correlator of two currents (one of which contains $\gamma_5$)
in a constant spatial vector field.
By virtue of the Ward identity, it is equal to the derivative
of the correlator without external field with respect to the spatial momentum
flowing along the light quark line.
This is minus derivative with respect to
the spatial momentum flowing along the heavy quark line.
But the HQET propagator does not depend on the spatial momentum,
and hence the derivative vanishes.
We have checked by a direct calculation, using the technique of~\cite{BG2},
that the two--loop correction (Fig.~\ref{Fmes}c) vanishes
in the case of naive anticommuting $\gamma_5$
(which is the correct one to use in the weak current).

The diagram with cut line (Fig.~\ref{Fmes}c) gives
\begin{equation}
\Pi_A(t) = \frac{f_\pi^2}{4} \left( 1
- \frac{i{<}\overline{q}q{>}t}{3f_\pi^2}
+ \frac{5}{36}m_1^2 t^2
- \frac{im_0^2{<}\overline{q}q{>}t^3}{48f_\pi^2}
\right)\,.
\label{OPEmes}
\end{equation}
Thus the appearance of $g_A$ of the constituent quark is entirely caused by
interaction with the quark condensate.
The first three terms in~(\ref{OPEmes}) agree with~\cite{Ovch,Ital,BBKR};
the fourth one is taken from~\cite{Ital}.
The correlator~(\ref{OPEmes}) corresponds to the spectral density
\begin{equation}
\rho_A(\omega) = \frac{f_\pi^2}{4} \left( \delta(\omega)
- \frac{{<}\overline{q}q{>}}{3f_\pi^2}\delta'(\omega)
- \frac{5}{36}m_1^2\delta''(\omega)
+ \frac{m_0^2{<}\overline{q}q{>}}{48f_\pi^2}\delta'''(\omega)
\right)\,.
\label{rho0}
\end{equation}

We have calculated the one--loop correction
to the leading dimension contribution.
Only the one--particle--irreducible diagram (Fig.~\ref{Fmes}e)
has to be calculated,
all the other contributions can be taken from~\cite{BG2}.
The correlator of bare currents in $d=4-2\varepsilon$ dimensions is
\begin{equation}
\Pi_A^0(\omega) = \frac{if^2}{4\omega} \left[ 1 +
\frac{C_F g_0^2 (-2\omega)^{-2\varepsilon}}{(4\pi)^{d/2}}
\frac{(d-2)(d-7)}{2(d-3)}\Gamma(-\varepsilon)\Gamma(1+2\varepsilon)
\right].
\label{bare}
\end{equation}
After renormalization of the heavy--light HQET currents,
it gives the finite result
\begin{eqnarray}
&&\Pi_A(\omega) = \frac{if_\pi^2}{4\omega} \left[1
+ C_F \frac{\alpha_s}{4\pi} \left(-6\log\frac{-2\omega}{\mu}+5\right)
\right],
\nonumber\\
&&\rho_A(\omega) = \frac{f_\pi^2}{4} \left[\delta(\omega)
- 6C_F\frac{\alpha_s}{4\pi}\frac{1}{\omega} \right].
\label{ren}
\end{eqnarray}
In order to obtain the coordinate--space result at $t=-i\tau$,
one has to calculate the exact $d$--dimensional spectral density,
take its Laplace transform, and go to the limit $d\to4$ only at the end:
\begin{equation}
\Pi_A(\tau) = \frac{f_\pi^2}{4} \left[1 + C_F \frac{\alpha_s}{4\pi}
\left(-6\log\frac{2}{\mu\tau}+6\gamma+5\right) \right],
\label{rad}
\end{equation}
where $\gamma$ is the Euler constant.
Using the technique of~\cite{BG2},
it is not difficult to calculate the two--loop correction to~(\ref{ren}).
However, we decided not to derive the next--to--next--to--leading term
in the sum rule for $g_1 F^2$, because the sum rule for $F^2$ is known
only with the next--to--leading accuracy~\cite{BG,BBBD}.

As usual, we assume that the excited states' contribution is dual to that
of the theoretical spectral density~(\ref{ren}) above the effective
continuum threshold $\varepsilon_c$. Equating the phenomenological
expression for the correlator to the theoretical one at an imaginary time
$t=-i/E$ we obtain the sum rule
\begin{eqnarray}
&&\left(\frac{g_1F^2(\mu)}{E}+c\right)e^{-\varepsilon/E}
\nonumber\\
&&\quad{} = \frac{f_\pi^2}{4}
\Biggl[1+C_F\frac{\alpha_s}{4\pi}\left(-6\log\frac{2\varepsilon_c}{\mu}
+6\E\left(\frac{\varepsilon_c}{E}\right)+5\right)
\label{SRmes}\\
&&\qquad{} - \frac{{<}\overline{q}q{>}}{3f_\pi^2E}
- \frac{5}{36}\frac{m_1^2}{E^2}
+ \frac{m_0^2{<}\overline{q}q{>}}{48f_\pi^2 E^3}
\Biggr]\,,
\nonumber
\end{eqnarray}
where
\begin{equation}
\E(x)=\int\limits_0^x\frac{1-e^{-t}}{t}dt=\left\{
\begin{array}{ll}
\frac{x}{1\cdot1!}-\frac{x^2}{2\cdot2!}+\frac{x^3}{3\cdot3!}-\cdots & x\ll1\\
\log x+\gamma+O(e^{-x}/x), & x\gg1
\end{array}
\right.
\label{E}
\end{equation}
The $\mu$--dependence of the right--hand side of~(\ref{SRmes}) coincides
with that of $F^2(\mu)$, with the accepted accuracy. 
The correction appears to be about $25-30\%$ at $\mu=1$~GeV 
and therfore is important.

\begin{figure}[ht]
\includegraphics[bb=90 590 380 770,clip,width=\linewidth]{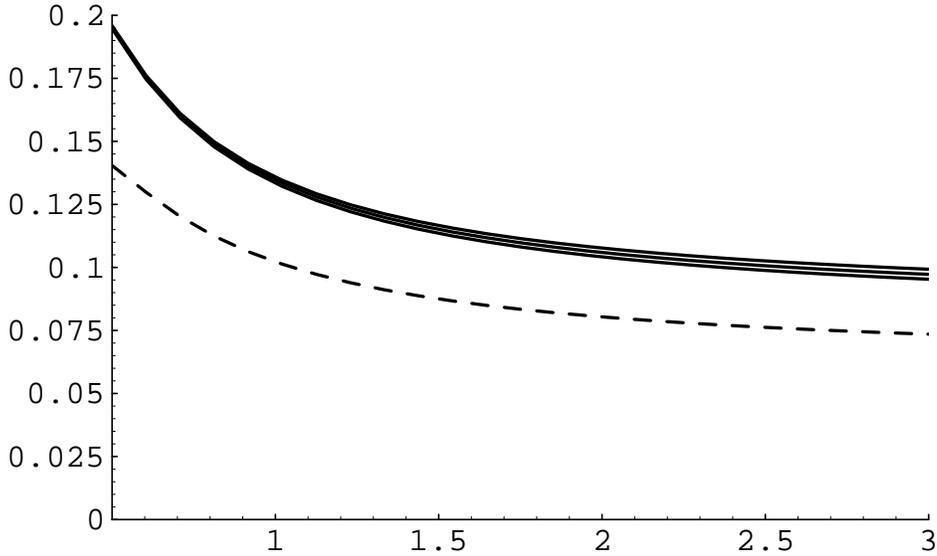}
\caption{$g_1 F^2/(380\text{MeV})^3$ as a function of Borel parameter $E$:
dashed line---without the perturbative correction,
solid lines---with the perturbative correction
($\varepsilon=500$MeV~\cite{BBBD}, $\mu=1$GeV, $\alpha_s=0.34$,
${<}\overline{q}q{>}=-(240\text{MeV})^3$,
continuum threshold $\varepsilon_c=1$GeV, $1.25$GeV, $1.5$GeV).}
\label{Fgmes}
\end{figure}

We can multiply~(\ref{SRmes}) by $\exp(\varepsilon/E)$
and differentiate in $E$ in order to exclude $c$.
This procedure is equivalent to finding a plato linear
in the Borel parameter~\cite{IS}.
The result is shown in Fig.~\ref{Fgmes}.
The value of $F$ is a subject of controversy in the current literature.
We choose the middle value of~\cite{BBBD} $F^2=(380\text{MeV})^3$
as a reference point; the sum rule produces $g_1 F^2$,
therefore lower values of $F^2$ yield larger values of $g_1$.
The OPE series~(\ref{SRmes}) behaves as
$1+270\text{MeV}/E-(170\text{MeV}/E)^2-(240\text{MeV}/E)^3$,
so it seems that the expansion is applicable at $E>500$MeV.
This is close to the lower bound of the applicability region
of the ordinary sum rule~\cite{Shuryak}.
The continuum contribution is small until very large values of $E$,
because the continuum only appears at the $\alpha_s/\pi$ level~(\ref{ren}).
At large $E$ the sum rule~(\ref{SRmes}) (without the perturbative correction)
leads to
\begin{equation}
g_1 F^2 = \tfrac{1}{4}f_\pi^2\varepsilon
- \tfrac{1}{12}{<}\overline{q}q{>}.
\label{Dualmes}
\end{equation}
At lower $E$, stability of this sum rule is not very good.
Our final numerical estimate is $g_1 F^2/(380\text{MeV})^3=0.1\div 0.2$.
The effect of the perturbative correction is moderate,
and dependence on the continuum threshold is very week.

The numerical results of the previous works differ from each other
largely due to different choices of the value of $F^2$.
Expressed in a common scale, they are remarkably similar:
$g_1 F^2/(380\text{MeV})^3=0.16\pm0.03$~\cite{Ovch},
$0.16\pm0.04$~\cite{Ital}, $0.17\pm0.01$~\cite{BBKR} 
(in the last result only diviation with Borel parameter is indicated and 20\% 
level of accuracy is assumed), $0.14\pm0.03$~\cite{CF}.
The perturbative correction was not included in earlier works,
and hence they had the continuum spectral density equal to zero.
Ovchinnikov~\cite{Ovch} included the first excited state
in the phenomenological side, assuming its energy known.
However, he only included $\delta(\omega-\varepsilon')$
in $\rho_A(\omega)$, and not $\delta'(\omega-\varepsilon')$.
We see no justification for such an assumption.
Taking $c_2\delta'(\omega-\varepsilon')$ into account would add
one more unknown parameter $c_2$ to the problem.

\section{Baryons}
\label{Bar}

Correlators of the $\Sigma\Sigma$ and $\Lambda\Sigma$ currents have the
$A$--terms
\begin{eqnarray}
&&i{<}T j^{ij}_{\Sigma l}(x) j^*_{\Sigma i'j'm}(0){>}_A =
i\varepsilon_{lmn}A^{(i}_{n(i'}\delta^{j)}_{j')}\delta(\vec{x})\Pi_A(x_0),
\nonumber\\
&&i{<}T j^{ij}_\Lambda(x) \vec{\jmath}\,^*_{\Sigma i'j'}(0){>}_A =
\vec{A}\,^{[i}_{(i'}\delta^{j]}_{j')}\delta(\vec{x})\Pi_A(x_0).
\label{Bcor}
\end{eqnarray}
If we define ${<}0|j^{ij}_\Lambda|\Lambda{>}=F_\Lambda e^{ij}_\Lambda$,
${<}0|\vec{\jmath}\,^{ij}_\Sigma|\Sigma{>}=F_\Sigma\vec{e}\,^{ij}_\Sigma$,
then the physical states' contributions to the correlators (Fig.~\ref{Fphys})
are
\begin{eqnarray}
\Pi_A(\omega)&=&\frac{2g_2F_\Sigma^2}{(\omega-\varepsilon_\Sigma)^2}
+\frac{c}{\omega-\varepsilon_\Sigma}+\cdots
\label{Bphys}\\
\Pi_A(\omega)&=&\frac{2g_3F_\Lambda F_\Sigma}
{(\omega-\varepsilon_\Lambda)(\omega-\varepsilon_\Sigma)}
+\frac{c_\Lambda}{\omega-\varepsilon_\Lambda}
+\frac{c_\Sigma}{\omega-\varepsilon_\Sigma}+\cdots
\nonumber
\end{eqnarray}
It is impossible to separate the $g_3$ term from the mixed
$\Lambda$--excited and $\Sigma$--excited contributions unambiguously.
We can do it approximately if
$\Delta=\varepsilon_\Sigma-\varepsilon_\Lambda\ll
\varepsilon_c-\varepsilon_{\Lambda,\Sigma}$
because in such a case partial fraction decomposition of the first term would
give large contributions to $c_{\Lambda,\Sigma}\sim1/\Delta$
with the opposite signs while the natural scale of $c_{\Lambda,\Sigma}$
in~(\ref{Bphys}) is
$c_{\Lambda,\Sigma}\sim 1/(\varepsilon_c-\varepsilon_{\Lambda,\Sigma})$.
This is not a defect of the sum rule but the uncertainty inherent to $g_3$
which can be defined only when $\Delta$ is small compared to
the chiral symmetry breaking scale.
We choose to require $c_\Lambda=c_\Sigma$; the choices $c_\Lambda=0$
or $c_\Sigma=0$ would be equally good.

\begin{figure}[p]
\includegraphics[width=\linewidth]{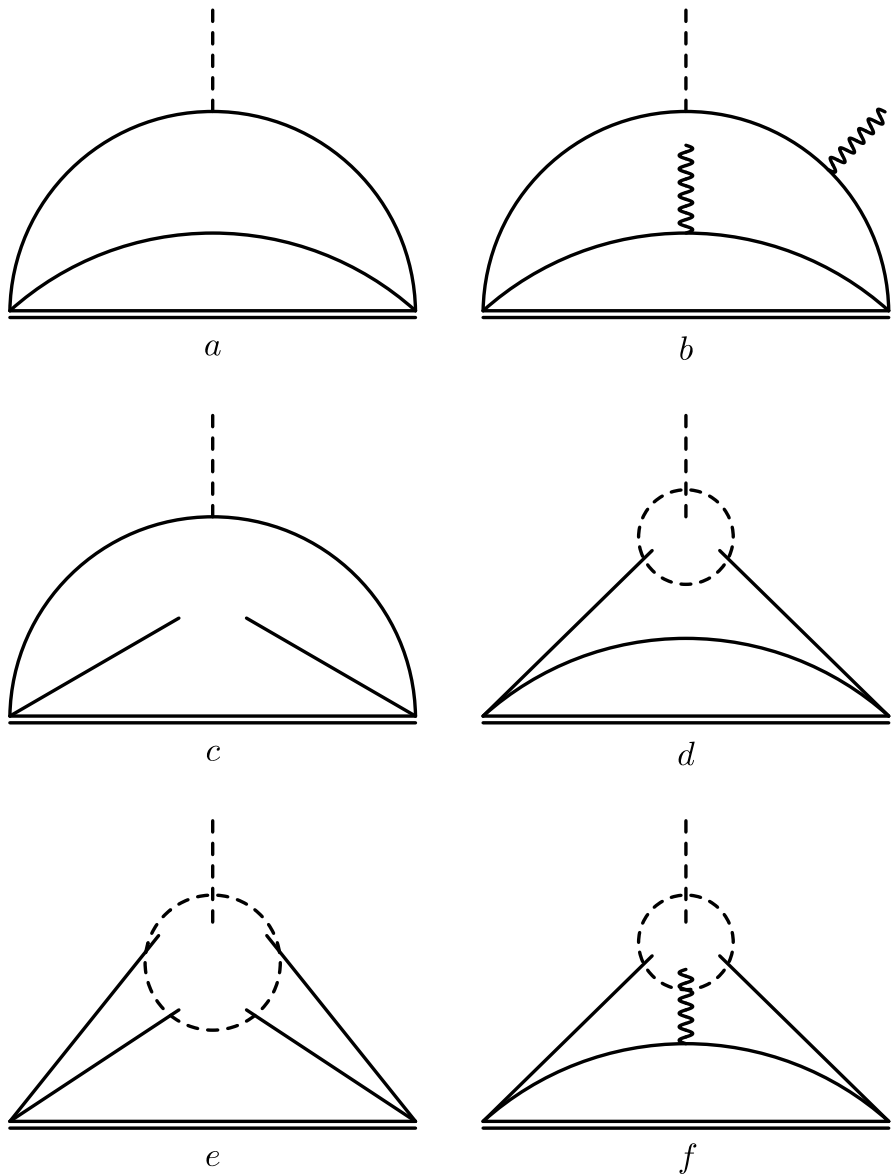}
\caption{Correlator of baryon currents.}
\label{Fbar}
\end{figure}

The baryonic correlators in the OPE framework are described by the diagrams
in Fig.~\ref{Fbar}.
The diagrams Fig.~\ref{Fbar}a--c with the non--cut quark
line interacting with the axial field vanish due to~(\ref{Prop}).
We use the factorization approximation for the four--quark condensate
in Fig.~\ref{Fbar}e.
In this approximation two diagonal correlators
${<}j_1j_1{>}$ and ${<}j_2j_2{>}$ coincide in both $\Sigma\Sigma$ and
$\Lambda\Sigma$ cases, as well as ${<}j_{\Lambda1}j_{\Sigma2}{>}$ and
${<}j_{\Lambda2}j_{\Sigma1}{>}$.
This is similar to the usual heavy baryon
sum rules~\cite{GY,GKY}, and confirms the observation that
$F_{\Lambda1}=F_{\Lambda2}$ and $F_{\Sigma1}=F_{\Sigma2}$ within the
factorization approximation to the sum rules.
Only even--dimensional condensates contribute to the diagonal $\Sigma\Sigma$
and the nondiagonal $\Lambda\Sigma$ correlators:
\begin{equation}
\Pi_A(t)=\frac{2N!\,f_\pi^2}{N\pi^2t^3}\left[1
+ \left(\frac{5}{3}\pm\frac{C_B}{C_F}\right)\frac{m_1^2t^2}{12}
+ \frac{\pi^2{<}\overline{q}q{>}^2t^4}{6Nf_\pi^2}\right],
\label{OPEbar1}
\end{equation}
where $C_B/C_F=1/(N-1)$ (this term comes from the diagram Fig.~\ref{Fbar}f).
Only odd--dimensional condensates contribute to the
nondiagonal $\Sigma\Sigma$ and the diagonal $\Lambda\Sigma$ correlators:
\begin{equation}
\Pi_A(t)=\frac{2N!\,{<}\overline{q}q{>}}{3N\pi^2t^2}\left[1-
\frac{3\pi^2f_\pi^2t^2}{2N}
\left(1+\tfrac{1}{16}m_0^2t^2+\tfrac{5}{36}m_1^2t^2\right)
\right].
\label{OPEbar2}
\end{equation}
These correlators correspond to the spectral densities
\begin{equation}
\rho_A(\omega)=\frac{N!\,f_\pi^2}{N\pi^2}\left[\omega^2
- \left(\frac53\pm\frac{C_B}{C_F}\right)\frac{m_1^2}{6}
\right],
\quad
\rho_A(\omega)=-\frac{2N!\,{<}\overline{q}q{>}\omega}{3N\pi^2}
\label{rho}
\end{equation}
(plus $\delta(\omega)$ and its derivatives).

We use the standard continuum model $\rho_A^{\text{cont}}(\omega)=
\rho_A^{\text{theor}}\vartheta(\omega-\varepsilon_c)$.
Equating the OPE~(\ref{OPEbar1}, \ref{OPEbar2}) and the spectral
representation at $t=-i/E$, we obtain the sum rules
\begin{eqnarray}\label{SRule}
&&\left(\frac{2g_2F_\Sigma^2}{E}+c\right)e^{-\varepsilon_\Sigma/E}
\nonumber\\
&&\quad=\frac{4f_\pi^2}{\pi^2}E^3\left[f_2\left(\varepsilon_c/E\right)
- \frac{13m_1^2}{72E^2}f_0\left(\varepsilon_c/E\right)
+ \frac{\pi^2{<}\overline{q}q{>}^2}{18f_\pi^2E^4}\right]\\
&&\quad=-\frac{4{<}\overline{q}q{>}}{\pi^2}E^2
\left[f_1\left(\varepsilon_c/E\right)
+ \frac{\pi^2f_\pi^2}{2E^2}
\left(1-\frac{m_0^2}{16E^2}-\frac{5m_1^2}{36E^2}\right)
\right],
\nonumber
\end{eqnarray}
\begin{eqnarray}\label{SRule1}
&&\left(\frac{2g_3F_\Lambda F_\Sigma}{\Delta}\tanh\frac{\Delta}{2E}
+\frac{c}{2}\right)
\left(e^{-\varepsilon_\Lambda/E}+e^{-\varepsilon_\Sigma/E}\right)
\label{SRbar} \nonumber \\
&&\quad=\frac{4f_\pi^2}{\pi^2}E^3\left[f_2\left(\varepsilon_c/E\right)
- \frac{7m_1^2}{72E^2}f_0\left(\varepsilon_c/E\right)
+ \frac{\pi^2{<}\overline{q}q{>}^2}{18f_\pi^2E^4}\right]\\
&&\quad=-\frac{4{<}\overline{q}q{>}}{\pi^2}E^2
\left[f_1\left(\varepsilon_c/E\right)
+ \frac{\pi^2f_\pi^2}{2E^2}
\left(1-\frac{m_0^2}{16E^2}-\frac{5m_1^2}{36E^2}\right)
\right],
\nonumber
\end{eqnarray}
where $f_n(x)=1-e^{-x}\sum_{m=0}^{n} x^m/m!$.

\begin{figure}[p]
\includegraphics[bb=90 590 380 770,clip,width=\linewidth]{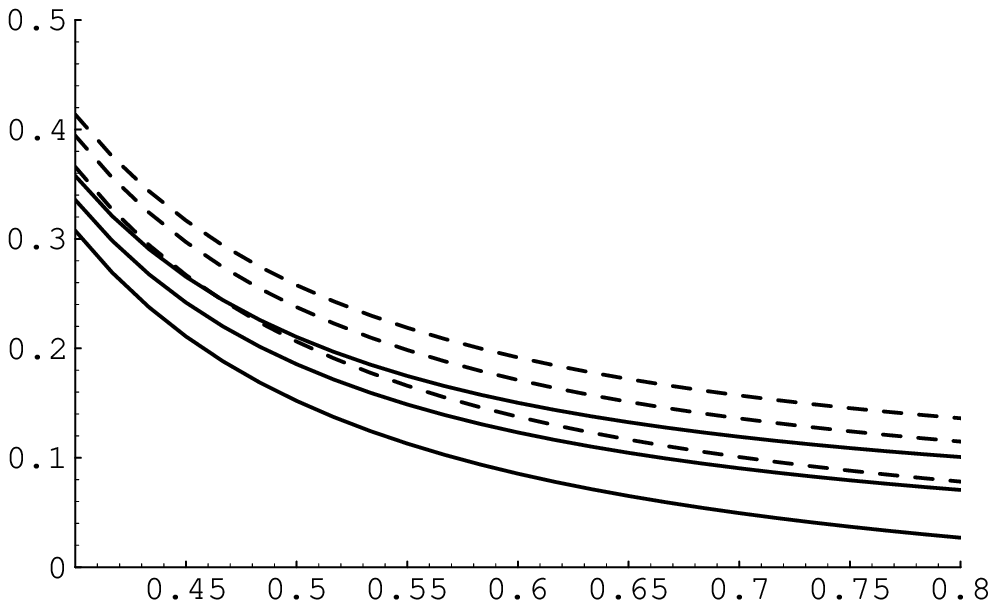}
\includegraphics[bb=90 590 380 770,clip,width=\linewidth]{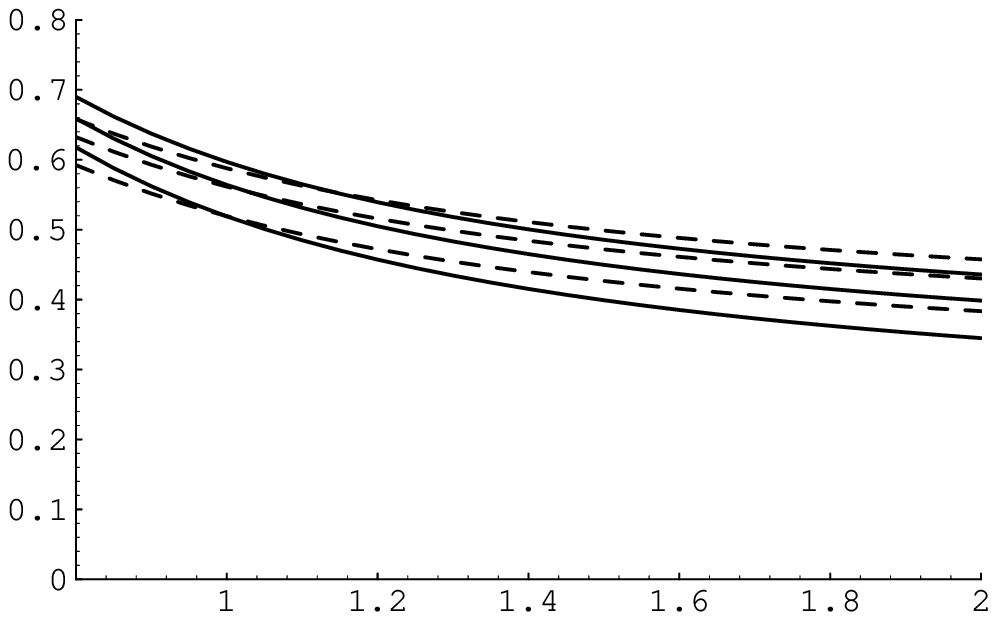}
\caption{Sum rules for $g_{2,3}$ (solid and dashed lines):
a---diagonal $\Sigma\Sigma$ and nondiagonal $\Lambda\Sigma$ sum rules;
b---nondiagonal $\Sigma\Sigma$ and diagonal $\Lambda\Sigma$ sum rules.}
\label{Fgbar}
\end{figure}

Now we can start the numerical analysis.
We adopt the energies and residues of $\Lambda$ and $\Sigma$
obtained from the HQET sum rules~\cite{GY,GKY}:
\begin{eqnarray*}
&& \varepsilon_\Lambda = 0.8\text{GeV}, \quad F_\Lambda = 0.03\text{GeV}^3, \\
&& \varepsilon_\Sigma  = 1.0\text{GeV}, \quad F_\Sigma  = 0.04\text{GeV}^3.
\end{eqnarray*}
We assume that the effective continuum threshold for the $A$--terms
of the $\Sigma$--$\Sigma$ correlators is the same as without the external
field: $\varepsilon_c=(1.3\pm0.1)$GeV; for the $\Lambda$--$\Sigma$
correlators we use $\varepsilon_c=(1.1\pm0.1)$GeV.

Fig.~\ref{Fgbar}a shows the results for 
$g_2 F_\Sigma^2/(1.6\cdot10^{-3}\text{GeV}^6)$ and
$g_3 F_\Lambda F_\Sigma/(1.2\cdot10^{-3}\text{GeV}^6)$ obtained from
the diagonal $\Sigma$--$\Sigma$ and the nondiagonal $\Lambda$--$\Sigma$ 
correlators (the first formulae in~(\ref{SRule}) and~(\ref{SRule1})).
The spectral density~(\ref{rho}) behaves like $\omega^2$,
and hence the continuum contribution to the sum rules~(\ref{SRbar})
quickly grows: it is equal to the result at $E\approx600$MeV
and is three times larger than the result at $E\approx900$MeV.
If we assume that the uncertainty in the continuum contribution
is, e.~g., 30\%, then we can't trust the sum rules at $E>900$MeV.
The lower bound of the applicability region is determined by
the convergence of the OPE series for the correlators.
It behaves like $1-(190\text{MeV}/E)^2+(245\text{MeV}/E)^4$
for the $\Sigma$--$\Sigma$ correlator;
in the $\Lambda$--$\Sigma$ case $140$MeV enters instead of $190$MeV.
It seems that OPE should be applicable at $E>400$MeV.
The sum rules in this window are unstable and therefore
yield no definite predictions.

Fig.~\ref{Fgbar}b shows the results for
$g_2 F_\Sigma^2/(1.6\cdot10^{-3}\text{GeV}^6)$ and
$g_3 F_\Lambda F_\Sigma/(1.2\cdot10^{-3}$GeV$^6$) obtained from
the nondiagonal $\Sigma$--$\Sigma$ and the diagonal $\Lambda$--$\Sigma$ 
correlators (the second formulae in~(\ref{SRule}) and~(\ref{SRule1})).
The spectral density~(\ref{rho}) behaves like $\omega$,
and the continuum contribution grows not so quickly:
it is equal to the result at $E\approx1$GeV
and is three times larger than the result at $E\approx1.6$GeV.
The OPE series behaves like
$1+(290\text{MeV}/E)^2\left[1-(150\text{MeV}/E)^2\right]$,
and the applicability region starts at a larger $E$.
Stability of these sum rules is quite good. 
The sum rules give $g_{2,3}=0.4\div 0.7$.

\textbf{In conclusion}, the meson sum rule in an external axial field,
with the perturbative correction and three successive power corrections,
gives the value $g_1 F^2/(380\text{MeV})^3=0.1\div0.2$,
which is much lower than the constituent quark model expectation $g_1=0.75$
unless a substantially lower value of $F^2$ is used.
The sum rules for $g_{2,3}$ following from nondiagonal 
$\Sigma$--$\Sigma$ and diagonal $\Lambda$--$\Sigma$ baryon correlators in
an external axial field suggest $g_{2,3}= 0.4\div 0.7$,
while diagonal $\Sigma$--$\Sigma$ and nondiagonal $\Lambda$--$\Sigma$ baryon
sum rules have too large uncertainties.

\textbf{Acknowledgements}. We are grateful to V.~M.~Belyaev, V.~M.~Braun,
V.~L.~Chernyak, and A.~Yu.~Khodjamirian for useful discussions.
A.~G.~G.'s work was supported in part by BMBF under contract
No.\ 06 MZ865; he is grateful to Institut f\"ur Physik, Mainz,
for hospitality during the final stage of the work.
O.~I.~Y.'s work was supported in part by BMBF under contract
No.\ 05 7WZ91P (0).

\end{document}